\documentclass[12pt]{iopart}

\usepackage{epsfig}

\begin{document}

\title{Type-II superconductivity in W$_{5}$Si$_{3}$-type Nb$_{5}$Sn$_{2}$Al}

\author{Jifeng Wu$^{1,2}$, Bin Liu$^{1,2}$, Yanwei Cui$^{2}$, Hangdong Wang$^{4}$, Zhicheng Wang$^{3}$, Zhi Ren$^{2}$\footnote[1]{Electronic address: zhi.ren@wias.org.cn}, Guanghan Cao$^{3}$}

\address{$^{1}$Department of Physics, Fudan University, Shanghai 200433, P. R. China}
\address{$^{2}$School of Science, Westlake Institute for Advanced Study, Westlake University, 18 Shilongshan Road, Hangzhou 310064, P. R. China}
\address{$^{3}$Department of Physics, Zhejiang University, Hangzhou 310027, P. R. China}
\address{$^{4}$Department of Physics, Hangzhou Normal University, Hangzhou 310036, P. R. China}

\date{\today}

\begin{abstract}
We report the discovery of superconductivity in the ternary aluminide Nb$_{5}$Sn$_{2}$Al, which crystallizes in the W$_{5}$Si$_{3}$-type structure with one-dimensional Nb chains along the $c$-axis.
It is found that the compound has a multiband nature and becomes a weakly coupled, type-II superconductor below 2.0 K.
The bulk nature of superconductivity is confirmed by the specific heat jump, whose temperature dependence shows apparent deviation from a single isotropic gap behavior.
The lower and upper critical fields are estimated to be 2.0 mT and 0.3 T, respectively.
From these values, we derive the penetration depth, coherence length and Ginzburg-Landau parameter to be 516 nm, 32.8 nm and 15.6, respectively.
By contrast, the isostructural compound Ti$_{5}$Sn$_{2}$Al dose not superconduct above 0.5 K.
A comparison of these results with other W$_{5}$Si$_{3}$-type superconductors suggests that $T_{\rm c}$ of these compounds correlates with the average number of valence electrons per atom.
\end{abstract}
\pacs{74.10.-v, 74.25.-q, 74.70.Dd}
\maketitle

\section{\label{sec:level1}Introduction}
Some structural types have been shown to favor the occurrence of superconductivity (SC) in intermetallic compounds \cite{SCreview}.
A well-known example is the cubic $A$15 materials with the general formula $A$$_{3}$$B$, in which nearly fifty  superconductors were found and the highest $T_{\rm c}$ reaches above 20 K \cite{A15review}.
Recently, the $A$$_{5}$$B$$_{3}$ compounds and their derivatives emerge as another fertile ground for SC \cite{Nb5Ga3,Nb5Ge3,Nb5Si3,Nb5SiB2,NbSn2Ga,Ta5Sn2Ga,W5Si3,Mo5SiB2,W5SiB2,Zr5Sb3,HfSbRu,ZrSbRu,Mo5PB2,Zr5Ge3}. These materials mainly crystallize in three different structural types, namely, tetragonal Cr$_{5}$B$_{3}$-type \cite{Cr5B3type}, W$_{5}$Si$_{3}$-type \cite{W5Si3type}, and hexagonal Mn$_{5}$Si$_{3}$-type \cite{Mn5Si3type}.
In particular, the W$_{5}$Si$_{3}$ and $A$15 types are similar in that they both consist of one dimensional chains of the $A$ atoms, which could give rise to a large density of states at the Fermi level [$N$($E_{\rm F}$)] \cite{largeNEF}.
So far, except for W$_{5}$Si$_{3}$ itself \cite{W5Si3}, superconductors of this structural type include Nb$_{5}$$X_{3}$ ($X$=Si, Ge and Ga) \cite{Nb5Ga3,Nb5Ge3,Nb5Si3}, Nb$_{5}$Sn$_{2}$Ga \cite{NbSn2Ga}, Ta$_{5}$Ga$_{2}$Sn \cite{Ta5Sn2Ga}, Hf$_{5}$Sb$_{2.5}$Ru$_{0.5}$ \cite{HfSbRu}, and Zr$_{5}$Sb$_{2.4}$Ru$_{0.6}$ \cite{ZrSbRu}.
The exploration and study of new member will help to identify the factors that govern $T_{\rm c}$ in these materials.

Ternary aluminides Nb$_{5}$Sn$_{2}$Al and Ti$_{5}$Sn$_{2}$Al were first synthesized in 1990s and found to adopt the W$_{5}$Si$_{3}$-type structure \cite{Nb521}.
However, to our knowledge, their physical properties have not been investigated to date.
In this paper, we present measurements of resistivity, Hall coefficient, magnetic suceptibility and specific heat on high-quality polycrystalline samples of Nb$_{5}$Sn$_{2}$Al and Ti$_{5}$Sn$_{2}$Al.
We show that Nb$_{5}$Sn$_{2}$Al is a type-II superconductor with $T_{\rm c}$ of 2.0 K, and the behavior of its specific heat jump is inconsistent with Bardeen-Cooper-Schreiffer (BCS) prediction for a single isotropic gap.
Nevertheless, the isostructural Ti$_{5}$Sn$_{2}$Al with remains normal down to 0.5 K, albeit with a larger Sommerfeld coefficient. We compare these results with other W$_{5}$Si$_{3}$-type superconductors and discuss the dependence of their $T_{\rm c}$ on the valence electron count.

\section{\label{sec:level1}Experimental}
Polycrystalline $M_{5}$Sn$_{2}$Al ($M$= Nb, Ti) samples were synthesized by using a solid state reaction method, similar to the previous report \cite{Nb521}.
High purity powders of $M$ (99.99\%), Sn (99.99\%) and Al (99.999\%) with the stoichiometric ratio of 5:2:1 were homogenized and pressed into pellets in an argon-filled glove-box.
The pellets were then loaded in an alumina crucible, which was sealed in a tantalum (Ta) tube.
The Ta tube was protected in a argon-filled sealed quartz tube, which was heated to 900-1000 $^{\rm o}$C in a muffle furnace.
After maintaining for 10 days, the quartz tube was quenched into cold water.
The reaction was repeated for several times with intermediate grinding to ensure sample homogeneity.

The purity of the resulting $M_{5}$Sn$_{2}$Al samples was checked by powder X-ray diffraction (XRD) using a PANalytical x-ray diffractometer with a monochromatic Cu-K$_{\alpha1}$ radiation at room temperature.
The chemical compositions of these samples were measured with an energy-dispersive X-ray (EDX) spectrometer (Model Octane Plus). The spectra were collected on different locations of each sample, and the Al content was not included in the analysis due to contribution from Al sample holders.
For consistency reason, all the physical property measurements were performed on samples derived from the same pellet.
A part of the pellet was cut into regular-shaped samples for transport and specific heat measurements, and the remaining part was crushed into powders.
The electrical resistivity was measured using a standard four-probe method.
Resistivity, Hall coefficient and specific heat measurements down to 1.8 K were performed on a Quantum Design PPMS-9 Dynacool.
Measurements of resistivity and specific heat down to 0.5 K were carried out on a Quantum Design PPMS-9 Evercool II.
The two sets of specific-heat data agree well within 5\% in the overlapped temperature range.
The dc magnetization was measured on crushed powders with a commercial SQUID magnetometer (Quantum Design MPMS3).

\section{\label{sec:level1}Results}

\subsection{\label{sec:level1}{Crystal Structure and chemical composition}}
The structure of $M$$_{5}$Sn$_{2}$Al is sketched in Fig. 1(a) and (b).
All the atoms are distributed orderly in the lattice, and there are two distinct crystallographic sites for the $M$ atoms, namely $M$(1) at 16$k$ site and $M$(2) at 4$b$ site.
Notably, these atoms form linear and zigzag chains along the $c$-axis, respectively.
For Nb$_{5}$Sn$_{2}$Al, this one dimensional arrangement of Nb atoms is reminiscent of that observed in Nb$_{3}$Sn.
Figure 1(c) shows the XRD patterns of the Nb$_{5}$Sn$_{2}$Al and Ti$_{5}$Sn$_{2}$Al samples at room temperature, together with structural refinement profiles using the GSAS program.
The observed and calculated diffraction patterns show a good agreement for both cases, demonstrating that they adopt the tetragonal W$_{5}$Si$_{3}$-type structure.
As can be seen in Table 1, the refined lattice parameters are $a$ = 10.638 {\AA}, $c$ = 5.225 {\AA} for Nb$_{5}$Sn$_{2}$Al and  $a$ = 10.558 {\AA}, $c$ = 5.262 {\AA} for Ti$_{5}$Sn$_{2}$Al, which are in good agreement with the previous report \cite{Nb521}.
It is also noted that, for Ti$_{5}$Sn$_{2}$Al, there exists additional diffraction peaks due to unknown impurities, whose maximum intensity is around one tenth that of the main phase.
This suggests that the impurity phases occupy $\sim$10\% of the sample volume.
On the other hand, EDX measurements yield the Nb:Sn molar ratio of 5:2.06(6) and Ti:Sn ratio of 5:2.09(3) for the Nb$_{5}$Sn$_{2}$Al and Ti$_{5}$Sn$_{2}$Al samples, respectively.
These results indicate that both compounds are stoichiometric within the experimental error.
In addition, a Ti:Sn ratio in the range of 5:2.55$-$2.80 is observed in a small part of the Ti$_{5}$Sn$_{2}$Al sample, which is presumably ascribed to the impurities detected in XRD.

\subsection{\label{sec:level1}{Resistivity and Hall coefficient}}
Figure 2(a) shows the temperature dependence of resistivity for Nb$_{5}$Sn$_{2}$Al and Ti$_{5}$Sn$_{2}$Al.
For both samples, the resistivity shows a weak metallic behavior with a concave curvature, similar to that observed in other $A$$_{5}$$B$$_{3}$ compounds \cite{Mo5SiB2,W5SiB2,Zr5Sb3,Mo5PB2}. Actually, the data of Nb$_{5}$Sn$_{2}$Al is an almost rigid upward shift of that of Ti$_{5}$Sn$_{2}$Al, pointing to essentially the same temperature-dependent electron scattering mechanism for the two samples.
However, their low temperature behavior is quite different.
As can be seen in the inset of Fig. 2(a), when cooling below 2.5 K, the resistivity of Nb$_{5}$Sn$_{2}$Al drops rapidly to zero, evidencing a superconducting transition.
By contrast, for Ti$_{5}$Sn$_{2}$Al, the decrease in resistivity tends to saturate and no SC is observed down to 0.5 K.

Figure 2(b) shows the temperature dependence of 1/(e$R_{\rm H}$) for the two samples, where $e$ is the electron charge and $R_{\rm H}$ is the low-field Hall coefficient.
The $R_{\rm H}$ of Nb$_{5}$Sn$_{2}$Al is negative at room temperature and becomes positive below 200 K. A similar sign reversal in $R_{\rm H}$ is observed for Ti$_{5}$Sn$_{2}$Al, but its $R_{\rm H}$ changes back to negative again at $\sim$100 K. These results provide clear evidence that these materials are multiband systems with both electrons and holes.
Nevertheless, at temperatures below 50 K, $R_{\rm H}$ is nearly temperature independent and the Hall resistivity is linear with magnetic field [see the inset of Fig. 2(b)] for both samples.
Hence it appears that, in this temperature range, their electrical transport is dominated by a single type of carriers.
Assuming a one-band model, the hole and electron densities are estimated to be 1.3 $\times$ 10$^{22}$ cm$^{-3}$ and 1.5 $\times$ 10$^{22}$ cm$^{-3}$ for Nb$_{5}$Sn$_{2}$Al and Ti$_{5}$Sn$_{2}$Al, respectively.

\subsection{\label{sec:level1}{Magnetic susceptibility}}
The occurrence of SC in Nb$_{5}$Sn$_{2}$Al is corroborated by the magnetic susceptibility results of powder samples shown in Fig. 3. Here the demagnetization effect is taken into consideration assuming that the powders consist of cubic-shaped grains with the demagnetization factor $N_{\rm d}$ = 0.3. A strong diamagnetic signal is observed in both field-cooling (FC) and zero-field cooling (ZFC) data.
In addition, a divergence is seen between the ZFC and FC curves, which is characteristic of a type-II superconducting behavior.
The linear extrapolation of the initial ZFC diamagnetic transition intersects with the baseline at 2.0 K (see the inset of Fig. 3), which coincides with the onset of zero resistivity.
In light of this coincidence, we determine $T_{\rm c}$ as 2.0 K for Nb$_{5}$Sn$_{2}$Al.
At 0.4 K, the ZFC and FC data correspond to shielding and Meissner fractions of 85\% and 22\%, respectively,
and hence clearly indicate bulk SC.

\subsection{\label{sec:level1}{Specific heat}}
Figure 4(a) shows the low temperature specific heat data for Nb$_{5}$Sn$_{2}$Al and Ti$_{5}$Sn$_{2}$Al.
A strong anomaly is observed in Nb$_{5}$Sn$_{2}$Al near its $T_{\rm c}$, which further confirms the bulk nature of SC.
On the other hand, its normal state data can be well described by the Debye model $C_{\rm p}$/$T$ = $\gamma$ + $\beta$$T^{2}$,
where $\gamma$ and $\beta$ are the Sommerfeld and phonon specific heat coefficients, respectively.
The fit yields $\gamma$ = 19.0 mJ mol$^{-1}$ K$^{-2}$ and $\beta$ = 0.396 mJ mol$^{-1}$ K$^{-4}$.
The Debye temperature $\Theta_{\rm D}$ is then calculated as $\Theta_{\rm D}$ = (12$\pi^{4}$$N$$R$/5$\beta$)$^{\frac{1}{3}}$ = 340 K,
where $N$ = 8 and $R$ is the molar gas constant 8.314 J mol$^{-1}$ K$^{-1}$.
Applying the same analysis to the data of Ti$_{5}$Sn$_{2}$Al gives $\gamma$ = 31.5 mJ mol$^{-1}$ K$^{-2}$, $\beta$ = 0.2709 mJ mol$^{-1}$ K$^{-4}$, and $\Theta_{\rm D}$ = 386 K.
Compared with Nb$_{5}$Sn$_{2}$Al, the larger $\Theta_{\rm D}$ of Ti$_{5}$Sn$_{2}$Al is as expected, but its larger $\gamma$ value is somewhat surprising.

Figure 4(b) shows the normalized electronic specific heat of Nb$_{5}$Sn$_{2}$Al at zero field after subtraction of the phonon contribution, plotted a function of reduced temperature $T$/$T_{\rm c}$.
One can see that the normalized specific heat jump $\Delta$$C_{\rm el}$/$\gamma$$T_{\rm c}$ $\sim$ 0.86 is considerably smaller than the BCS value of 1.43 \cite{BCStheory}.
Furthermore, the temperature dependence of the specific heat jump shows a significant deviation from the BCS theory, especially in the low temperature region.
Note that an extrapolation of $C_{\rm el}$/$\gamma$$T$ to 0 K yields a negative value. This, together with the absence of additional XRD peaks from impurities [see Fig. 1(c)], suggests that the sample has a superconducting volume fraction close to 100\%.
It therefore appears that the small specific heat jump is of intrinsic origin, which points to the presence of multiple gaps as in MgB$_{2}$ \cite{MgB2} and Lu$_{2}$Fe$_{3}$Si$_{5}$ \cite{LuFeSi}, or even gap nodes.
Nevertheless, the $C_{\rm el}$/$\gamma$$T$ data at 0.5 K is still as large as 50\% of its normal-state value, and hence to distinguish these scenarios, $C_{\rm p}$/$T$ measurements down to mK range would be needed.

Assuming that SC in Nb$_{5}$Sn$_{2}$Al is mediated by electron-phonon coupling, the corresponding constant, $\lambda_{\rm ph}$, can be estimated by using the inverted McMillan formula \cite{Mcmillan},
\begin{equation}
\lambda_{\rm ph} = \frac{1.04 + \mu^{\ast} \rm ln(\Theta_{\rm D}/1.45\emph{T}_{\rm c})}{(1 - 0.62\mu^{\ast})\rm ln(\Theta_{\rm D}/1.45\emph{T}_{\rm c}) - 1.04},
\end{equation}
where $\mu^{\ast}$ is the Coulomb repulsion pseudopotential.
Taking an empirical $\mu^{\ast}$ value of 0.13,  we obtain $\lambda_{\rm ph}$ = 0.46, which implies weak coupling SC in Nb$_{5}$Sn$_{2}$Al.
This is consistent with the low value of $\Delta$$C_{\rm el}$/$\gamma$$T_{\rm c}$.

\subsection{\label{sec:level1}{Critical fields and superconducting parameters}}
Figure 5(a) shows the isothermal magnetization curves at various temperatures for Nb$_{5}$Sn$_{2}$Al.
The effective lower critical field $B_{\rm c1}^{\ast}$ at each temperature is defined as the value at which the
curve deviates from its initial linear behavior. The resulting $B_{\rm c1}^{\ast}$ is plotted as a function of temperature in Fig. 5(b), and the extrapolation of the data to zero temperature yields $B_{\rm c1}^{\ast}$(0) = 1.4 mT
\cite{Bc1fit}.
After correction for the demagnetization factor, we obtain the zero-temperature lower critical field $B_{\rm c1}$(0) = $B_{\rm c1}^{\ast}$(0)/(1 $-$ $N_{\rm d}$) = 2.0 mT. Figure 5(c) shows the temperature dependence of resistivity for Nb$_{5}$Sn$_{2}$Al under various magnetic fields.
With increasing field, the resistive transition is gradually suppressed to lower temperatures, as expected.
Following the same criterion as shown above (see Subsection 3.2.), we determine $T_{\rm c}$ for each field as the temperature at which zero resistivity is achieved, and plot the data in Fig. 5(d).
Extrapolation of the upper critical field $B_{\rm c2}$ to zero temperature, using the Wathamer-Helfand-Hohenberg (WHH) theory \cite{WHH}, gives $B_{\rm c2}$(0) = 0.3 T.

With $B_{\rm c1}$(0) and $B_{\rm c2}$(0) determined, we can estimate various superconducting parameters of Nb$_{5}$Sn$_{2}$Al.
The Ginzburg-Landau (GL) coherence length $\xi_{\rm GL}$(0) as $\xi_{\rm GL}$(0) = $\sqrt{\Phi_{0}/2\pi B_{\rm c2}(0)}$ = 32.8 nm, where $\Phi_{0}$ = 2.07 $\times$ 10$^{-15}$ Wb is the flux quantum.
Then from the equations $B_{\rm c1}$(0)/$B_{\rm c2}$(0) = (ln$\kappa_{\rm GL}$ + 0.5)/(2$\kappa_{\rm GL}^{2}$) \cite{kappa} and $B_{\rm c1}$(0) = ($\Phi_{0}$/4$\pi$$\lambda_{\rm eff}^{2}$)(ln$\kappa_{\rm GL}$ + 0.5),
$\kappa_{\rm GL}$ = 15.6 and $\lambda_{\rm eff}$ = 516 nm are obtained.

\section{\label{sec:level1}Discussion}
Now let us compare the physical properties of isostructural Nb$_{5}$Sn$_{2}$Al and Ti$_{5}$Sn$_{2}$Al, which are summarized in Table 2.
As noted above, the $\gamma$ value of Ti$_{5}$Sn$_{2}$Al is $\sim$60\% larger than that of Nb$_{5}$Sn$_{2}$Al.
This means that Ti$_{5}$Sn$_{2}$Al has a considerably higher $N$($E_{\rm F}$), yet no SC is observed.
To gain more insight, we plot the $T_{\rm c}$ against the average number of valence electrons per atom ratio ($e$/$a$) for all known "W$_{5}$Si$_{3}$"-type superconductors.
According to the empirical rule proposed by Matthias \cite{MTrule}, $T_{\rm c}$ of intermetallic superconductors is expected to exhibit two maxima at $e$/$a$ close to 5 and 7 \cite{MTrule}.
Indeed, as can be seen from the figure, most of the "W$_{5}$Si$_{3}$"-type superconductors have $e$/$a$ in the range between 4.25 and 4.625,
although $T_{\rm c}$ is generally higher for superconductors containing group IVB elements than those containing group VB elements. Furthermore, in the latter case, a $T_{\rm c}$ maximum is found at $e$/$a$ $\sim$ 4.5, which is slightly lower than the empirical value of 4.7 \cite{MTrule}. These results suggest that this family of superconductors follows the Matthias rule and their $T_{\rm c}$ is mainly controlled by the band filling rather than $N$($E_{\rm F}$).
In this respect, the absence of SC in Ti$_{5}$Sn$_{2}$Al can be explained since its $e$/$a$ of 3.875 appears to fall below the lower bound for SC.
On the other hand, the $e$/$a$ of W$_{5}$Si$_{3}$ exceeds the above range and reaches 5.25, hinting at the existence of another $T_{\rm c}$ maximum at higher valence electron account.

Finally, we note that, despite similar Nb atom chains, the $T_{\rm c}$ of Nb$_{5}$Sn$_{2}$Al is nearly one order of magnitude smaller than that
of Nb$_{3}$Sn.
This corroborates the previous notion that the presence of these one-dimensional chains is not a sufficient condition for achieving high $T_{\rm c}$ \cite{Nb5Ge3,Nb5Si3}.                                                                  Instead, it is pointed out that the $\lambda_{\rm ph}$ value of Nb$_{5}$Sn$_{2}$Al is only about one-third that of Nb$_{3}$Sn \cite{eph},                                                                                         which highlights the importance of electron phonon interaction in enhancing $T_{\rm c}$.

\section{\label{sec:level1}Conclusion}
In summary, we have studied the transport, magnetic and thermodynamic properties of isostructural Nb$_{5}$Sn$_{2}$Al and Ti$_{5}$Sn$_{2}$Al.
The results demonstrate that Nb$_{5}$Sn$_{2}$Al is a type-II superconductor with a bulk $T_{\rm c}$ of 2.0 K and a non-BCS gap, while Ti$_{5}$Sn$_{2}$Al remains metallic down to 0.5 K.
Surprisingly, compared with Nb$_{5}$Sn$_{2}$Al, Ti$_{5}$Sn$_{2}$Al has a even larger Sommerfeld coefficient, suggesting that the absence of SC is not due to a reduction in $N$($E_{\rm F}$).
Nevertheless, these results are in line with the empirical rule based on the average number of valence electrons per atom, and at the value of $\sim$4.5 a $T_{\rm c}$ maximum is found for the "W$_{5}$Si$_{3}$"-type superconductors.
Further studies to increase the number of valence electrons and electron phonon coupling, which may result in higher $T_{\rm c}$, are highly desirable in future.

\section*{Acknowledgments}
This work is supported by the National Key Research and Development Program of China (No.2017YFA0303002) and the Fundamental Research Funds for the Central Universities of China.

\pagebreak[4]

\begin{figure}
\centering
\includegraphics[width=12cm]{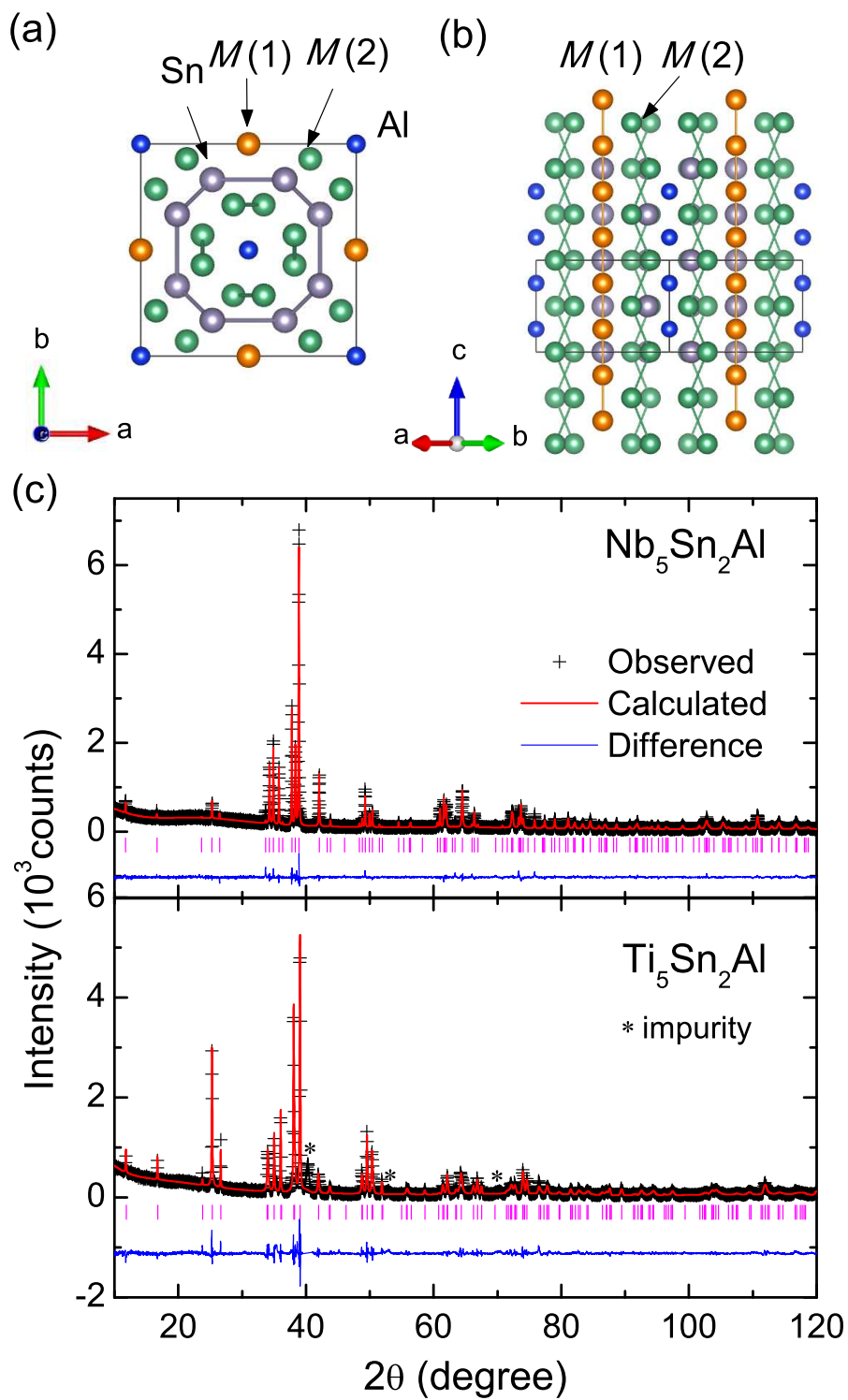}
\caption{\label{Fig.1}  Schematic structure of $M_{5}$Sn$_{2}$Al projected perpendicular (a) and parallel (b) to the $M$ atom chains.
(c) Powder x-ray diffraction patterns and their refinement profiles at room temperature for Nb$_{5}$Sn$_{2}$Al and Ti$_{5}$Sn$_{2}$Al.
The asterisks mark the impurity phases in the Ti$_{5}$Sn$_{2}$Al sample.}
\end{figure}

\begin{figure}[h]
\centering
\includegraphics[width=12cm]{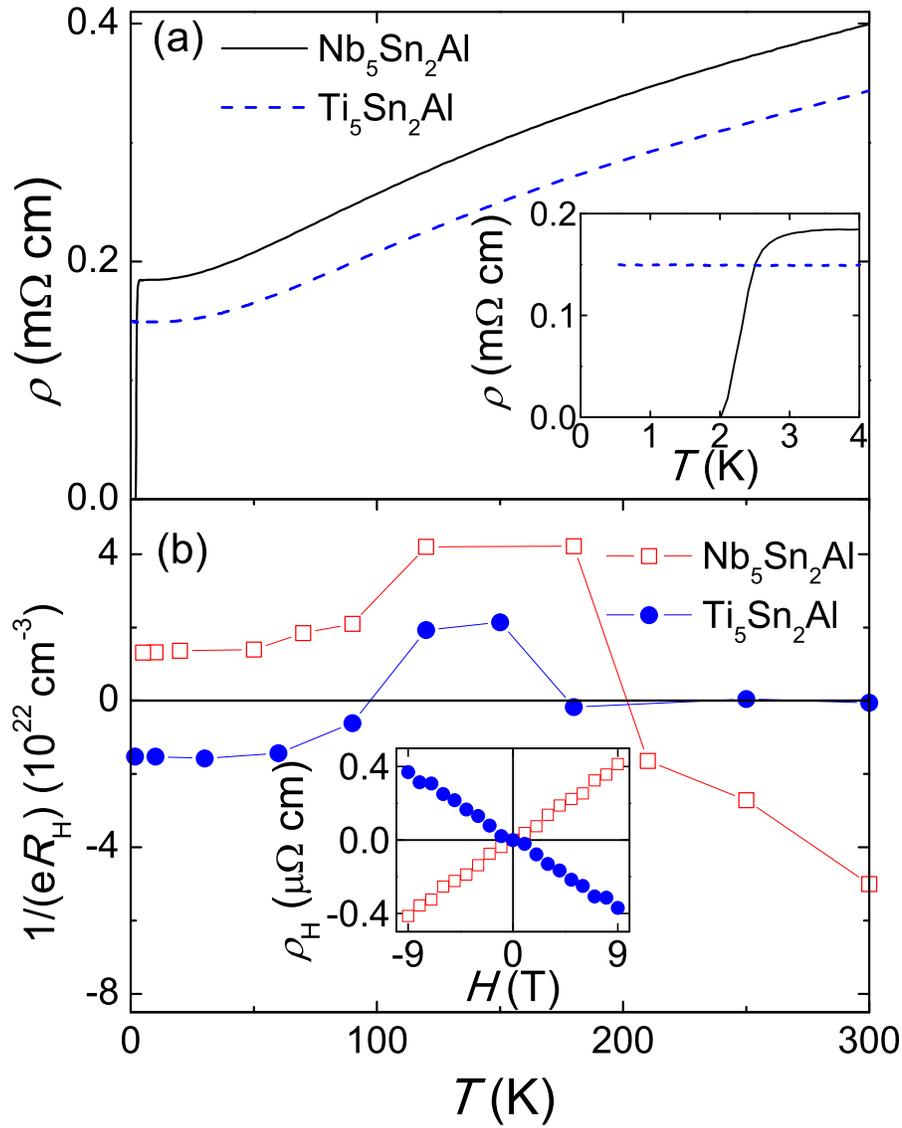}
\caption{\label{label} (a) Temperature dependence of resistivity for Nb$_{5}$Sn$_{2}$Al (solid line) and Ti$_{5}$Sn$_{2}$Al (dashed line).
The inset shows a zoom of the low temperature data below 4 K.
(b) Temperature dependence of inverse Hall coefficient for the two samples.
The inset shows the field dependence of Hall resistivity for these samples at 10 K.}
\end{figure}

\begin{figure}[h]
\centering
\includegraphics[width=12cm]{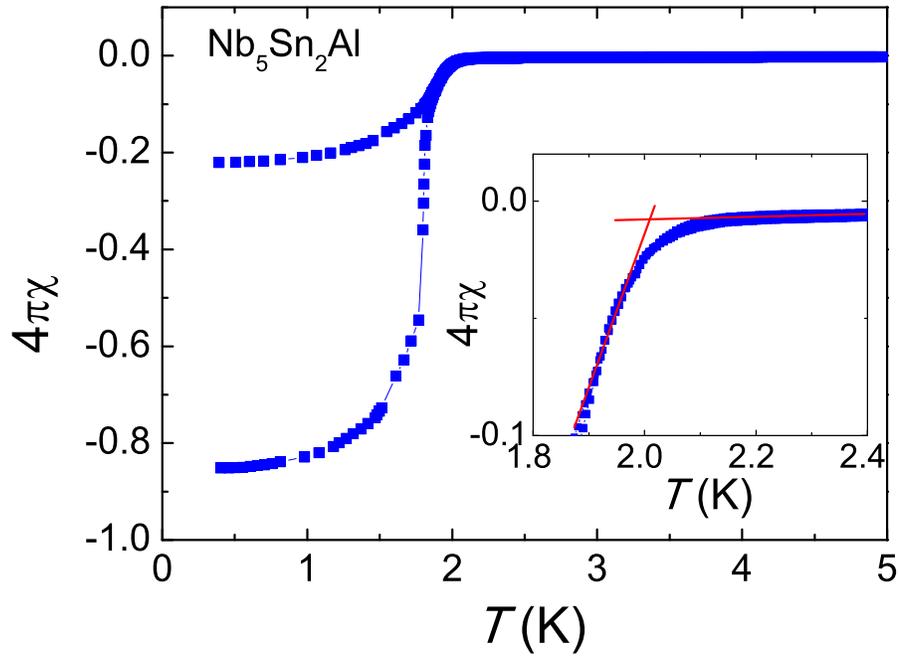}
\caption{\label{label} Temperature dependence of magnetic susceptibility below 5 K measured under an applied field of 0.45 mT.
The inset shows a zoom of the data between 1.8 and 2.4 K. The two solid lines are a guide to the eyes.}
\end{figure}

\begin{figure}[h]
\centering
\includegraphics[width=12cm]{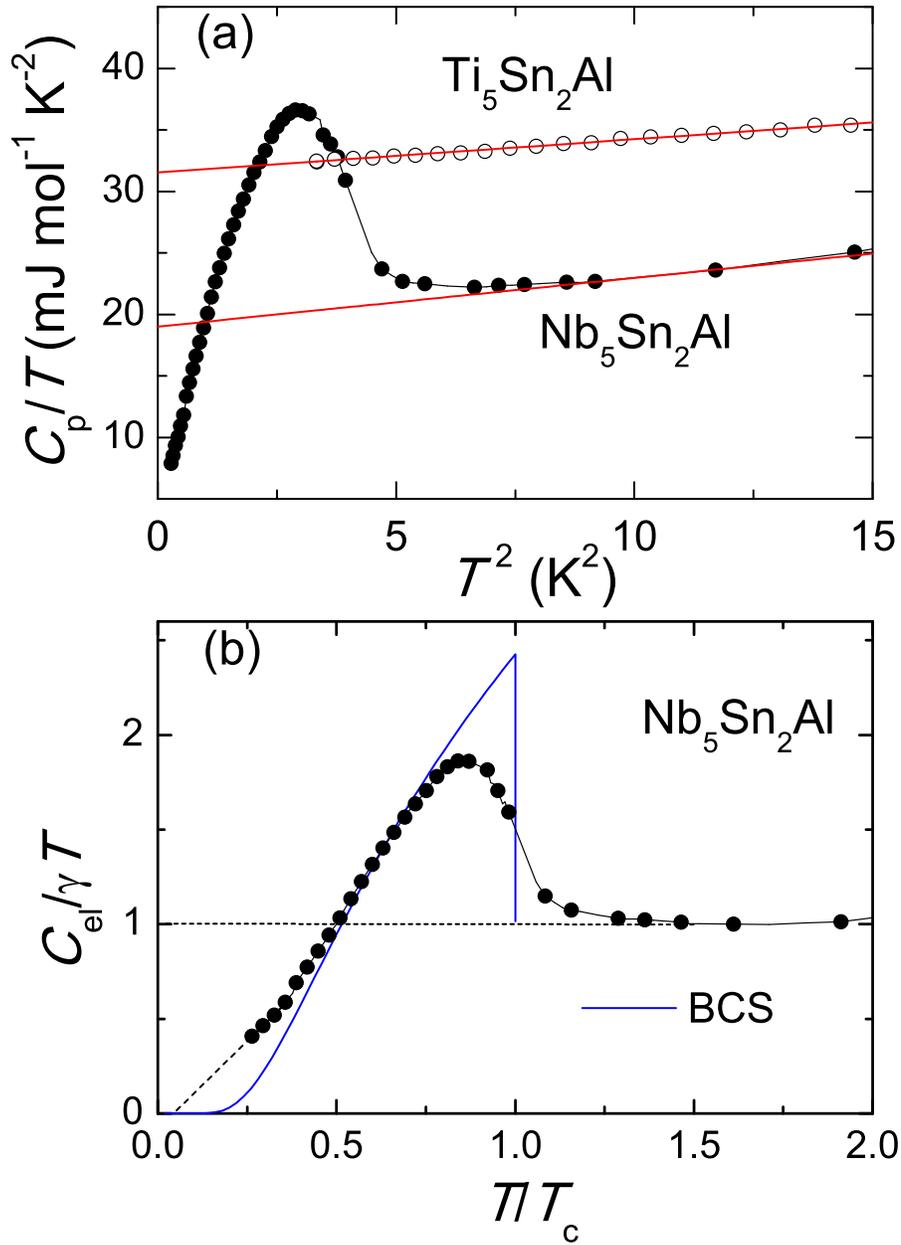}
\caption{\label{label} (a) Low temperature specific-heat data of Nb$_{5}$Sn$_{2}$Al and Ti$_{5}$Sn$_{2}$Al plotted as $C_{\rm p}/T$ versus $T^{2}$.
The solid lines are fits by the Debye model to the data (see text).
(b) Normalized electronic specific heat for Nb$_{5}$Sn$_{2}$Al.
The solid line denotes the theoretical BCS curve. The dashed lines are a guide to the eyes.}
\end{figure}

\begin{figure}[h]
\centering
\includegraphics[width=12cm]{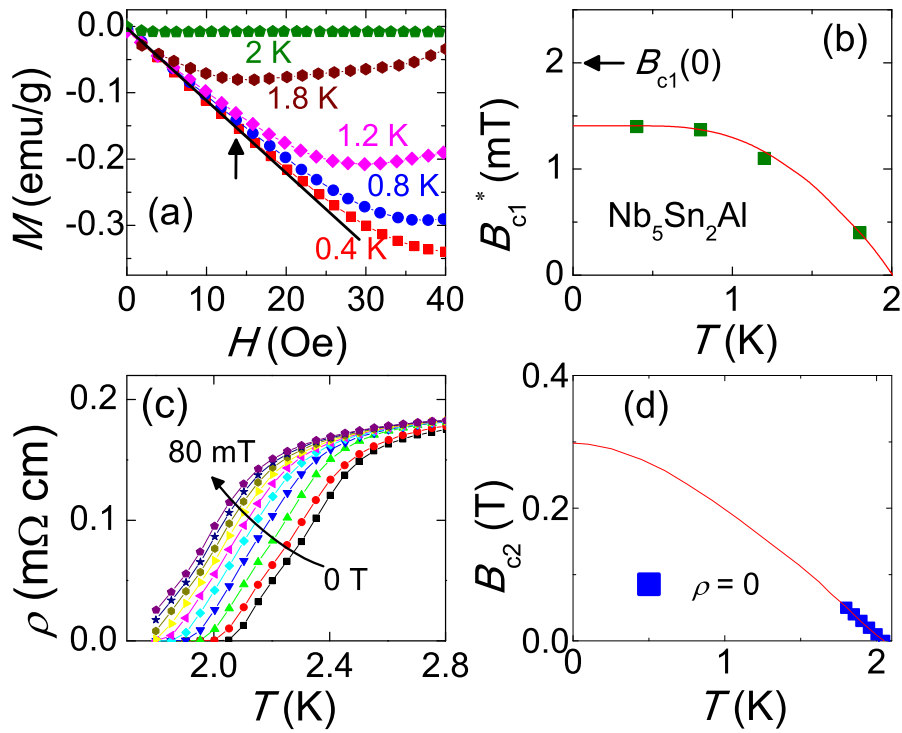}
\caption{\label{label} (a) Field dependence of magnetization curves for the Nb$_{5}$Sn$_{2}$Al sample measured after zero-field cooling to various temperatures.
The solid line denotes the initial linear behavior and the arrow marks the deviation from this linearity at 0.4 K.
(b) Temperature dependence of the effective lower critical field $B_{\rm c1}^{\ast}$.
The solid line is a fit to the data using the local dirty limit formula. The arrow marks the zero-temperature upper critical field $B_{\rm c1}$(0) after correction for the demagnetization factor.
(c) Temperature dependence of resistivity for the Nb$_{5}$Sn$_{2}$Al sample under various magnetic fields up to 80 mT in increment of 10 mT.
(d) The upper critical field $B_{\rm c2}$ plotted as a function of temperature for Nb$_{5}$Sn$_{2}$Al. The solid line is a WHH fit to the data.}
\end{figure}

\begin{figure}[h]
\centering
\includegraphics[width=12cm]{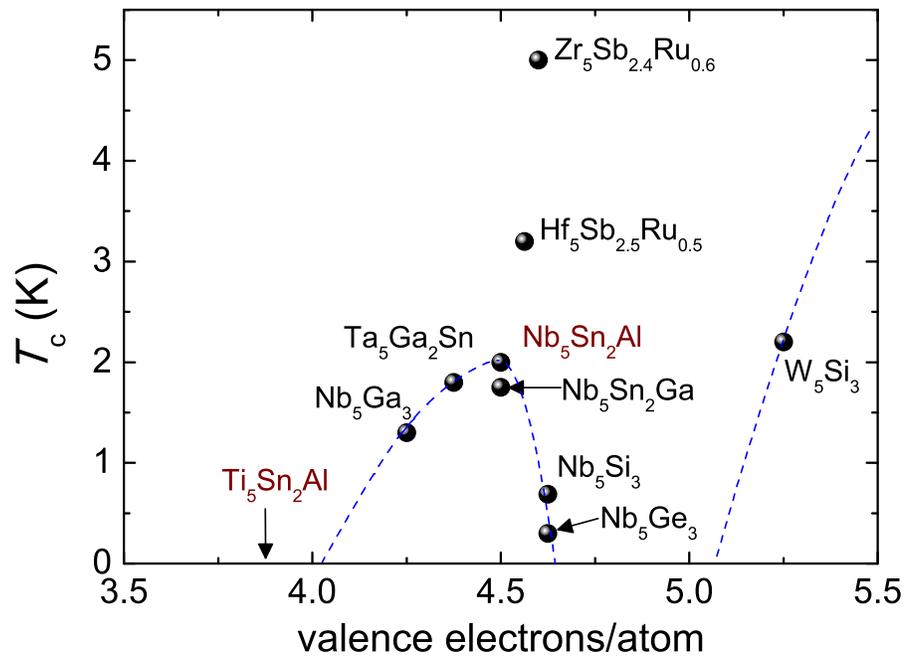}
\caption{\label{label} The superconducting transition temperature $T_{\rm c}$ plotted as a function of the average number of valence electrons per atom ratio for all known "W$_{5}$Si$_{3}$"-type superconductors.
The dashed lines are a guide to the eyes.}
\end{figure}

\clearpage
\begin{table}
\caption{Refined structural parameters of $M$$_{5}$Sn$_{2}$Al ($M$ = Nb, Ti). Here $a$ and $c$ are the lattice parameters, $V$ the unit cell volume, $R_{\rm wp}$ and $R_{\rm p}$ the weighted and unweighted $R$ factors.}

\begin{indented}
\lineup
\item[]\begin{tabular}{@{}*{5}{l}}
\br
\0\0& &Nb$_{5}$Sn$_{2}$Al & Ti$_{5}$Sn$_{2}$Al&\\
\0\0& Space group							 &  \multicolumn{2}{c}{$I$4/$mcm$}		\\
\0\0&$a$ ({\AA})				& 10.638(1)	& 10.558(2) \\
\0\0&$c$ ({\AA})				& 5.225(1)	& 5.262(1) \\
\0\0&$V$  ({\AA}$^{3}$)				& 591.46	& 586.57 \\
\0\0&$R_{\rm wp}$ 			 & 	  8.1\%	& 12.9\%	\\
\0\0&$R_{\rm p}$  				 & 	  5.7\%	&  9.6\%	\\
\mr
\0 Atoms& \0\0\0 site& \0\0 $x$ & \0\0 $y$ &\0\0 $z$\\
\0\0 $M$1&  \0\0\0 16k& 0.07308& 0.2134 &\0\0 0\\
\0\0 $M$2& \0\0\0 4b&\0\0 0 &\0 0.5 &\0 0.25\\
\0\0 Sn& \0\0\0 8h& 0.1683 & 0.6683 &\0\0 0\\
\0\0 Al& \0\0\0 4a&\0\0 0 &\0\0 0 &\0 0.25\\
\br
\end{tabular}
\end{indented}
\end{table}

\clearpage
\begin{table}
\caption{Physical parameters of Nb$_{5}$Sn$_{2}$Al and Ti$_{5}$Sn$_{2}$Al. Here $T_{\rm c}$ is the superconducting transition temperature, $B_{\rm c1}$(0) and $B_{\rm c2}$(0) the zero-temperature lower and upper critical fields for type-II superconductors,
$\lambda_{\rm eff}$ the effective penetration depth, $\xi_{\rm GL}$(0) the Ginzburg-Landau coherence length, $\kappa_{\rm GL}$ the Ginzburg-Landau parameter, $\Theta_{\rm D}$ the Debye temperature, $\gamma$ the Sommerfeld coefficient, $\lambda_{\rm ph}$ the electron-phonon coupling constant.}

\begin{indented}
\lineup
\item[]\begin{tabular}{@{}*{4}{l}}
\br
\0\0Parameters&Nb$_{5}$Sn$_{2}$Al &\0\0\0Ti$_{5}$Sn$_{2}$Al\\
\mr
\0\0$T_{\rm c}$ (K)							 & \0\0 2.0  & \0\0\0\0\0\0 $-$		\\
\0\0$B_{\rm c1}$(0)  (mT)				&\0\0 2.0	& \0\0\0\0\0\0 $-$ \\
\0\0$B_{\rm c2}$(0)  (T)				&\0\0 0.3	& \0\0\0\0\0\0 $-$ \\
\0\0$\lambda_{\rm eff}$  (nm)			 &\0\0 	  516	& \0\0\0\0\0\0 $-$	\\
\0\0$\xi_{\rm GL}$(0)  (nm)				 &\0\0 	  32.8	& \0\0\0\0\0\0 $-$	\\
\0\0$\kappa_{\rm GL}$			& \0\0	 15.6	& \0\0\0\0\0\0 $-$	\\
\0\0$\Theta_{\rm D}$ (K)				 & \0\0			340 	& \0\0\0\0\0 386		 \\
\0\0$\gamma$ (mJ mol$^{-1}$ K$^{-2}$)	& \0\0  19.0  & \0\0\0\0\0 31.5 \\
\0\0$\lambda_{\rm ph}$	 &\0\0   0.46 & \0\0\0\0\0\0 $-$  \\
\br
\end{tabular}
\end{indented}
\end{table}


\section*{References}
\begin{thebibliography}{00}
\bibitem{SCreview}
Matthias B T, Geballe T H, and Compton V B 1963 {\it Rev. Mod. Phys.} {\bf 35}, 1.

\bibitem{A15review}
Dew-Hughes D 1975 {\it Cryogenics} {\bf 15}, 435.

\bibitem{Nb5Ga3}
Havinga E E, Van Maaren M H, and Damsma H 1969 {\it Phys. Lett. A} {\bf 29}, 109.

\bibitem{Nb5Ge3}
Claeson T, Ivarsson J, and Rasmussen S E 1977 {\it J. Appl. Phys.} {\bf 48}, 3998.

\bibitem{Nb5Si3}
Willis J O and Waterstrat R M 1979 {\it J. Appl. Phys.} {\bf 50}, 2863.

\bibitem{Nb5SiB2}
Pesliev P, Gyurov G, and Soyanchev R 1986 {\it Izv. Khim.} {\bf 19}, 267.

\bibitem{NbSn2Ga}
Shishido T, Ukei K, Toyota N, Sasaki T, Watanabe Y, Motai K, Fukuda T, Takeya H, and Takei H 1989 {\it J. Cryst. Growth} {\bf 96}, 1.

\bibitem{Ta5Sn2Ga}
Shishido T, Ye J H, Toyota N, Ukei K, Sasaki T, Horiuchi H, and Fukuda T 1989 {\it Jpn. J. Appl. Phys.} {\bf 28}, 1519.

\bibitem{W5Si3}
Kawashima K, Muranaka T, Kousaka Y, Akutagawa S, and Akimitsu J 2009 {\it J. Phys.: Conf. Ser.} {\bf 150}, 052106.

\bibitem{Mo5SiB2}
Machado A J S, Costa A M S, Nunes C A, dos Santos C A M, Grant T, and Fisk Z 2011 {\it Solid State Commun.} {\bf 151}, 1455.

\bibitem{W5SiB2}
Fukuma M, Kawashima K, Maruyama M, and Akimitsu J 2011 {\it J. Phys. Soc. Jpn.} {\bf 80}, 024702.

\bibitem{Zr5Sb3}
Lv B, Zhu X Y, Lorenz B, Wei F Y, Xue Y Y, Yin Z P, Kotliar G, and Chu C W 2013 {\it Phys. Rev. B} {\bf 88}, 134520.

\bibitem{HfSbRu}
Xie W W, Luo X H, Seibel E, Nielsen M, and Cava R 2015 {\it Chem. Mater.} {\bf 27}, 4511.

\bibitem{ZrSbRu}
Xie W W, Luo X H, Phelan B F, and Cava R J 2015 {\it J. Mater. Chem. C} {\bf 3}, 8235.

\bibitem{Mo5PB2}
McGuire M A and Parker D S 2016 {\it Phys. Rev. B} {\bf 93}, 064507.

\bibitem{Zr5Ge3}
Li S, Liu X Y, Anand V, and Lv B 2018 {\it New J. Phys.} {\bf 20}, 013009.

\bibitem{Cr5B3type}
Bottcher P, Doert Th., Druska Ch., and Bradtmoller S 1997 {\it J. Alloy Compd.} {\bf 246}, 209.

\bibitem{W5Si3type}
Aronsson B 1955 {\it Acta Chem. Scand.} {\bf 9}, 1107.

\bibitem{Mn5Si3type}
Garcia E and Corbett J D 1988 {\it J. Solid State Chem.} {\bf 73}, 440.

\bibitem{largeNEF}
Paduani C 2009 {\it Solid State Commun.} {\bf 149}, 1269.

\bibitem{Nb521}
Pietzk M A and Schuster J C 1995 {\it J. Alloy Compd.} {\bf 230}, L10.

\bibitem{BCStheory}
Bardeen J, Cooper L N, and Schreiffer J R 1957 {\it Phys. Rev.} {\bf 108}, 1175.

\bibitem{MgB2}
Bouquet F, Fisher R A, Phillips N E, Hinks D G, and Jorgensen J D 2001 {\it Phys. Rev. Lett.} {\bf 87}, 047001.

\bibitem{LuFeSi}
Nakajima Y, Nakagawa T, Tamegai T, and Harima H 2008 {\it Phys. Rev. Lett.} {\bf 100}, 157001.

\bibitem{Mcmillan}
McMillan W L 1968 {\it Phys. Rev.} {\bf 167}, 331.

\bibitem{Bc1fit}
M. Tinkham 1975 {\it Introduction to superconductivity} (McGraw-Hill, New York).

\bibitem{WHH}
Werthamer N R, Helfand E, and Hohenberg P C 1966 {\it Phys. Rev.} {\bf 147}, 295.

\bibitem{kappa}
Hu C R 1972 {\it Phys. Rev. B} {\bf 6}, 1756.

\bibitem{MTrule}
Matthias B T 1955 {\it Phys. Rev.} {\bf 97}, 74.

\bibitem{eph}
Tutuncu H M, Srivastava G P, Bagci S, and Duman S 2006 {\it Phys. Rev. B} {\bf 74}, 212506.

\end{thebibliography}
\end{document}